\documentclass[twocolumn,preprintnumbers,aps,showpacs,superscriptaddress,prx]{revtex4-1}

\usepackage{bm}
\usepackage{amsmath}
\usepackage{color}
\usepackage{epsfig}
\usepackage{float}
\usepackage[pdftex, pdfstartview = {FitH},colorlinks=true,linkcolor=blue,citecolor=blue,urlcolor=black]{hyperref}
\usepackage[toc,page]{appendix}

\begin{document}

\title{Exact-exchange density functional theory of the integer quantum Hall effect: strict 2D limit}

%

\author{D. Miravet}
\email{dmiravet@gmail.com}
\affiliation{Centro At\'omico Bariloche, CNEA, CONICET, 8400 S. C. de Bariloche, R\'io Negro, Argentina}
\author{C. R. Proetto}
\email{proetto@cab.cnea.gov.ar}
\affiliation{Centro At\'omico Bariloche and Instituto Balseiro, 8400 S. C. de Bariloche, R\'io Negro, Argentina}

\begin{abstract}
A strict bidimensional (strict-2D) exact-exchange (EE) formalism within the framework of density-functional theory
(DFT) has been developed and applied to the case of an electron gas subjected to a strong perpendicular magnetic field,
that drives the system to the regime of the integer quantum Hall effect (IQHE). As the filling of the emerging Landau
levels proceeds, two main features results: {\it i)} the EE energy minimizes with a discontinuous derivative at every
integer filling factor $\nu$; and {\it ii)} the EE potential display sharp discontinuities at every integer $\nu$.
The present contribution provides a natural improvement as compared with the widely used local-spin-density
approximation (LSDA), since the EE energy functional fully contains the effect of the magnetic field, and includes
an inter-layer exchange coupling for multilayer systems. As a consistency
test, the LSDA is derived as the leading term of a low-field expansion of the EE energy and potential.

\end{abstract}
%

\maketitle
\section{Introduction}
Both the integer quantum Hall effect (IQHE) \cite{WvK11} and the fractional quantum Hall effect (FQHE) \cite{dSP97} are 
striking manifestations
of quantum mechanics in solid-state physics. However, besides their apparent similarity, important differences exist
between the two phenomena from the physical point of view. The standard view of the IQHE is a strict-2D single-particle
scenario dominated by the sequential filling of Landau levels (LL), plus a phenomenological description of the localization
effects induced by disorder \cite{GV05}. The FQHE, on the other side, is dominated by many-body effects among electrons inside a 
given LL.

In a previous work \cite{MFP17}, we have shown however that even in the IQHE the exchange interaction modify considerably the 
standard single-particle description, using as a theoretical tool a density functional theory (DFT) \cite{PY89} and an 
exact-exchange (EE) formalism \cite{G00}, applied to a quasi-2D electron gas localized in a finite-width semiconductor quantum well.
In a more general context, the description of interacting many-electron systems in external magnetic fields
in the framework of the optimized effective potential (OEP) method has been a long research goal of Hardy Gross and
coworkers  \cite{HKPRG08}. From the experimental side, evidence has been also found on the relevance of exchange
effects in the IQHE \cite{BKMPLMM11}.

The aim of this work is to proceed from these previous general quasi-2D situation towards an strict-2D limit,
where electrons are confined to move in a plane. The motivations for doing this are several: i) most of the models
of the IQHE and the FQHE use this dimensional approximation; ii) a considerable simplicity in the calculations is
achieved in the strict-2D limit; and iii) it is a realistic limit, in the sense that when only one-subband of the 
quasi-2D electron gas is occupied, the differences between the quasi-2D and the strict-2D limit are small, and can
be included by using the ``form factors''  \cite{B88}. By imposing on our general formalism the constraint that electrons
are confined to a plane, we may compare our results with one of the standard approximations for the study of the IQHE,
as is the strict-2D Local Spin Density Approximation (LSDA)  \cite{AS17}. The strict-2D LSDA has been also employed
in the past for the study of the FQHE, using {\it ad-hoc} generated exchange-correlation energy 
functionals  \cite{FV95}.

Proceeding in this way the solution of the complicated integral equation for the EE potential may be found analytically, 
shedding light over some subtle points such as the presence of discontinuities at every integer filling factor 
$\nu$  \cite{note}.
Also, the simplicity of our approach allow us to analyze in detail the low magnetic-field or large filling factor limit,
and obtain the strict-2D LSDA as the leading contribution both to the EE energy and potential. The present work also
provides naturally the EE potential on neighboring 2D layers (bilayers, trilayers, ...), solving in this way one of the 
major drawbacks of the strict-2D LSDA when applied to a multilayer situation, that is the absence of exchange interlayer
interactions. This problem has been already discussed in the literature, and solved using a variational Hartree-Fock 
theoretical approach  \cite{HM96}, which is outside the present DFT framework.

The remainder of this work is organized as follows: in Section \ref{secc:2}  we give a short review of our EE quasi-2D formalism
in the IQHE regime and explain how to constraint it to the strict-2D case. In Section \ref{secc:3} we provide the main results
and the associated discussions, and Section \ref{secc:conclu} is devoted to the conclusions. In Appendix \ref{appendix:limits} we give some details
about how to expand our results when the filling factor is much larger than one. 

\section{EE for quasi-2D electron systems: general formalism and strict-2D limit}
\label{secc:2}
\subsection{General formalism}
The electronic exchange energy can be defined as 
\begin{eqnarray}
 E_{\text{x}} &=& - \sum_{a,b,\sigma} f_a^{\sigma} f_b^{\sigma} \nonumber \\ 
 &\times& \int d^{3}{r} \int d^{3}{r}'
 \frac{\Psi_{a}^{\sigma}(\mathbf{r})^*\Psi_{b}^{\sigma}(\mathbf{r}')^*
 \Psi_{b}^{\sigma}(\mathbf{r})\Psi_{a}^{\sigma}(\mathbf{r}')}{2|\mathbf{r}-\mathbf{r}'|},
 \label{eq:Exact_exchange_energy}
\end{eqnarray}
in effective atomic units (length in units of the effective Bohr radius $a_0^* = \epsilon \hbar^2/e^2 m^*$,
and energy in units of the effective Hartree $Ha^* = m^* e^4 / \epsilon^2 \hbar^2$) \cite{units}. With $f_a^{\sigma},f_b^{\sigma}$ being 
the state $(a,b)$ and spin $(\sigma)$ finite temperature weights, taking values between 0 and 1.
For a quasi-2D electron gas (q2DEG) in the $x-y$ plane with an applied magnetic field along the $z$-direction, 
the wave
function can be written as 
$\Psi_{i,n,k}^{\sigma}(\mathbf{r}) =\psi_{n,k}(\bm{\rho})\lambda_{i}^{\sigma}(z)= 
\phi_{n}(x)\frac{e^{iky}}{\sqrt{L_y}}\lambda_{i}^{\sigma}(z)$
(in Landau gauge), where 
\begin{equation}
 \phi_{n}(x) =
 \frac{\exp\left[-\frac{\left(x-l_{B}^{2}k\right)^{2}}{2l_{B}^{2}}\right]}
 {\left[\sqrt{\pi}\, l_{B}\,2^{n}\left(n!\right)\right]^{1/2}}H_{n}\left(\frac{x-l_{B}^{2}k}{l_{B}}\right).
\end{equation}
Here $H_n(x)$ are the $n$-th Hermite polynomials, and $n \; (= 0,1,2,...)$ is the orbital quantum number
index. $k$ is the one-dimensional wave-vector label that distinguishes states within a given LL, each with a degeneracy $N_\phi = AB/\Phi_0$. $A$ is the area of the q2DEG in the $x-y$ plane, $B$ is the magnetic field strength in the $z$ direction, and $\Phi_0= ch/e$ is the magnetic flux number.   $l_B = \sqrt{c\hbar/eB}/a_0^*$ is the magnetic length. The $\lambda_i^{\sigma}(z)$ are the self-consistent KS eigenfunctions for electrons in subband $i \; (= 1,2,...)$, spin $\sigma \; (=\uparrow,\downarrow)$  and eigenvalue $\gamma_i^\sigma(\nu)$.
Substituting the last expression for the wave function in Eq. (\ref{eq:Exact_exchange_energy})
we obtain 
\begin{align}
 E_{\text{x}}= & -\frac{N_{\phi}}{2l_{B}}\sum_{i,j,\sigma}\int dz\int dz'\;\lambda_{i}^{\sigma}(z)\lambda_{i}^{\sigma}(z')\lambda_{j}^{\sigma}(z)\lambda_{j}^{\sigma}(z')
 \nonumber \\
 & \times\sum_{n,m}n_{i,n,\sigma}^{\text{2D}}n_{j,m,\sigma}^{\text{2D}}I_{n}^{m}(z-z').
 \label{eq:exact_exchange_energy_quasi-2D}
\end{align}
Here $n_{i,n,\sigma}^{\text{2D}}=\int g(\varepsilon-\varepsilon_{i,n}^{\sigma})f_{\text{FD}}(\varepsilon)d\varepsilon$
is the effective occupation factor of a LL labeled by \{$i,n,\sigma$\}. $f_{\text{FD}}(\varepsilon)=[1+e^{(\varepsilon-\mu)/(k_{B}T)}]^{-1}$
is the Fermi-Dirac distribution function, and $\mu$ is the chemical
potential. $g(\varepsilon)$ is the DOS normalized to $1$, so that $0\leq n_{i,n,\sigma}^{\text{2D}}\leq1$.
Under these conditions, $\sum_{i,n,\sigma}n_{i,n,\sigma}^{\text{2D}} = \nu = N / N_\phi$
is constant and defines $\mu$, with $N$ being the total number of electrons. Finally, 

\begin{align}
I_{n}^{m}(z-z') & =\frac{n_{<}!}{n_{>}!}\int_{0}^{\infty}dx\; e^{-x^{2}/2}({x^{2}}/{2})^{n_{>}-n_{<}}\nonumber \\
 & \times\left[L_{n_<}^{n_{>}-n_{<}}({x^{2}}/{2})\right]^{2}e^{-{x|z-z'|}/{l_{B}}},
 \label{I_nm}
\end{align}
as given elsewhere   \cite{MP16}. Here $L_n^m(x)$ are the generalized Laguerre polynomials, 
and $n_< = \min(n,m)$, $n_> = \max(n,m)$. The full 3D eigenvalues associated with $\Psi_{i,n,k}^{\sigma}(\mathbf{r})$
are given by 
$\varepsilon_{i,n}^{\sigma}(\nu) = \gamma_i^{\sigma}(\nu) + (n+1/2)\hbar \omega_c/Ha^* - |g|\mu_B B s(\sigma)/(2Ha^*)$.
Here $\omega_c = eB/m^*c$ is the cyclotron frequency, and the last term is the Zeeman splitting, with $s(\uparrow)=+1$,
and $s(\downarrow)=-1$.
In passing from Eq. (\ref{eq:Exact_exchange_energy}) to Eq. (\ref{eq:exact_exchange_energy_quasi-2D}), the sums
over partially filled LL have been treated as explained in Eq. (\ref{average}) below.
When the system has only one
subband occupied ($i=j=1$) Eq. (\ref{eq:exact_exchange_energy_quasi-2D}) simplifies
to 
\begin{align}
 E_{\text{x}}= & -\frac{N_{\phi}}{2l_{B}}\sum_{\sigma}\int dz\int dz'\rho_{\sigma}(z)\rho_{\sigma}(z')\nonumber \\
 & \times\frac{S_{1}^{\nu_{\sigma}}(|z-z'|)}{(\nu_{\sigma}N_{\phi})^{2}} \; ,
 \label{Ex}
\end{align}
where 
\begin{equation}
 S_{1}^{\nu_{\sigma}}(t)=\sum_{n,m}n_{1,n,\sigma}^{\text{2D}}n_{1,m,\sigma}^{\text{2D}}I_{n}^{m}(t) \; ,
\end{equation}
and $\rho_{\sigma}(z)=N_{\sigma} |\lambda_1^{\sigma}(z)|^2$, $\nu=\nu_{\uparrow}+\nu_{\downarrow}$.
The spin-dependent EE potential can be obtained from $v_{\text{x}}^{\sigma}(z)={\delta E_{\text{x}}}/{\delta\rho_{\sigma}(z)}$,
which reads
\begin{align}
 v_{\text{x}}^{\sigma}(z)=\frac{-1}{l_{B}N_{\phi}\nu_{\sigma}^{2}}\int dz'\rho_{\sigma}(z')S_{1}^{\nu_{\sigma}}(z-z')
 \nonumber \\
 -\frac{1}{2l_{B}N_{\phi}}\int dz\int dz'\rho_{\sigma}(z)\rho_{\sigma}(z')\nonumber \\
 \times\frac{\partial\left(S_{1}^{\nu_{\sigma}}(z-z')/\nu_{\sigma}^{2}\right)}{\partial\gamma_{1}^{\sigma}}
 \frac{\partial\gamma_{1}^{\sigma}}{\partial\rho_{\sigma}(z)} \; ,
 \label{inter}
\end{align}
where the first (second) term comes from the explicit (implicit) dependence
of $E_{\text{x}}$ on $\rho_{\sigma}(z)$. After some calculations, Eq. (\ref{inter})
becomes 

\begin{equation}
v_{\text{x}}^{\sigma}(z)=u_{\text{x}}^{\sigma}(z)+\overline{\Delta v}_{\text{x}}^{\;\sigma}\;,\label{eq:V_x}
\end{equation}
with $u_{\text{x}}^{\sigma}(z)$ being the first term in Eq. (\ref{inter}),
$\overline{\Delta v}_{\text{x}}^{\;\sigma}=\eta_{\text{x}}^{\nu_{\sigma}}-\bar{u}_{\text{x}}^{\sigma}$,
$\bar{u}_{\text{x}}^{\sigma}=\int\lambda_{1}^{\sigma}(z)^{2}{u}_{\text{x}}^{\sigma}(z)dz$,
and $\eta_{\text{x}}^{\nu_{\sigma}}=-\left\langle \rho_{\sigma}|S_{2}^{\nu_{\sigma}}|\rho_{\sigma}\right\rangle /\left(\nu_{\sigma}^{2}(N_{\phi})^{2}l_{B}\right)$. 

It remains to define

\begin{equation}
S_{2}^{\nu_{\sigma}}(t)=\frac{\sum_{n,m}\left({\partial n_{n,\sigma}^{\text{2D}}}/{\partial\gamma_{1}^{\sigma}}\right)n_{m,\sigma}^{2D}\; I_{n}^{m}(t)}{\sum_{n}\left({\partial n_{n,\sigma}^{\text{2D}}}/{\partial\gamma_{1}^{\sigma}}\right)}\;.\label{eq:S2}
\end{equation}

The expressions for $E_{\text{x}}$ and $v_{\text{x}}^{\sigma}(z)$
may be further simplified if we consider the low-temperature limit $T\rightarrow 0$ and suppose that the LL broadening 
$\Gamma$
is smaller than the energy difference between consecutive LL's with
the same spin ($\hbar\omega_{c}>\Gamma \gg k_B T$). Then, denoting
by $[\nu_{\sigma}]$ the integer part of $\nu_{\sigma}$, the occupation
factors are just given by 

\begin{align}
 n_{1,n,\sigma}^{\text{2D}}\equiv n_{n,\sigma}^{\text{2D}}= & \begin{cases}
 1 & n<\left[\nu_{\sigma}\right]\;,\\
 p_{\sigma} & n=\left[\nu_{\sigma}\right]\;,\\
 0 & n>\left[\nu_{\sigma}\right]\;,
\end{cases}\label{of}
\end{align}
where $p_{\sigma}=\nu_{\sigma}-\left[\nu_{\sigma}\right]$, and $0\leq p_{\sigma}<1$
is the fractional occupation factor of the more energetic occupied
LL with spin $\sigma$. This allow us to simplify the sums $S_{1}^{\nu_{\sigma}}(t)$
and $S_{2}^{\nu_{\sigma}}(t)$ as follows,
\begin{align}
 S_{1}^{\nu_{\sigma}}(t)= & \sum_{n_{i},n_{j}}n_{i,n_{i},\sigma}^{2D}n_{j,n_{j},\sigma}^{2D}I_{n_{i}}^{n_{j}}(t)\nonumber \; ,\\
 \simeq & \sum_{n_{i},n_{j}=0}^{[\nu_{\sigma}]-1}I_{n_{i}}^{n_{j}}(t)+
 2p_{\sigma}\sum_{n_{i}=0}^{[\nu_{\sigma}]-1}I_{n_{i}}^{[\nu_{\sigma}]}(t)+
 p_{\sigma}^{2}I_{[\nu_{\sigma}]}^{[\nu_{\sigma}]}(t).
 \label{DefS1}
\end{align}
To rewrite expression (\ref{eq:S2}) we use that 
$\frac{\partial n_{1,n_{j},\sigma}^{2D}}{\partial\varepsilon_{1}^{\sigma}} \simeq
\delta_{n_{j},[\nu_{\sigma}]}\times\frac{\partial n_{1,[\nu_{\sigma}],\sigma}^{2D}}{\partial\varepsilon_{1}^{\sigma}}$,
then 
\begin{align}
 S_{2}^{\nu_{\sigma}}(t)= 
 & \frac{1}{\sum_{n_{i}}\frac{\partial n_{1,n_{i},\sigma}^{2D}}{\partial\varepsilon_{1}^{\sigma}}}
 \sum_{n_{i},n_{j}}n_{i,n_{i},\sigma}^{2D}\frac{\partial n_{1,n_{j},\sigma}^{2D}}{\partial\varepsilon_{1}^{\sigma}}
 I_{n_{i}}^{n_{j}}(t)\nonumber \; , \\
 \simeq & \sum_{n_{i}}n_{i,n_{i},\sigma}^{2D}I_{n_{i}}^{[\nu_{\sigma}]}(t)\nonumber \; , \\
 = & \sum_{n_{i}=0}^{[\nu_{\sigma}]-1}I_{n_{i}}^{[\nu_{\sigma}]}(t)+p_{\sigma}I_{[\nu_{\sigma}]}^{[\nu_{\sigma}]}(t).
 \label{DefS2}
\end{align}
In the last line of Eqs. (\ref{DefS1}) and (\ref{DefS2}), it must be fulfilled the constraint $[\nu_{\sigma}]-1 \ge 0$.
For the special case $[\nu_{\sigma}]=0$, $S_{1}^{\nu_{\sigma}}(t) = p_{\sigma}^{2}I_{0}^{0}(t)$,
and $S_{2}^{\nu_{\sigma}}(t) = p_{\sigma}I_{0}^{0}(t)$.

\subsection{Strict-2D limit}

Up to this point we have followed the same calculation scheme as in the Ref.   \cite{MFP17}.
Now we will focus on the strict-2D limit of the q2DEG, that can be obtained using
the replacement $|\lambda_{1}^{\sigma}(z)|^{2}\rightarrow\delta(z)$
in the previous expressions for the EE energy and potential in Eqs. (\ref{Ex}) and (\ref{inter}), obtaining respectively

\begin{eqnarray}
 e_{\text{x}}(r_s,\nu) \equiv \frac{E_{\text{x}}}{N}
 &=&-\frac{1}{\sqrt{2} \; r_s \; \nu^{3/2}} \sum_{\sigma}S_{1}^{\nu_{\sigma}}(0) \nonumber \;, \\ 
 &=& e_{\text{x}}^{\uparrow}(r_s,\nu) + e_{\text{x}}^{\downarrow}(r_s,\nu) \; ,
 \label{eq:Ex_2D_frac_fill_factor}
\end{eqnarray}
and
\begin{equation}
 v_{\text{x}}^{\sigma}(z) = \frac{-1}{r_s \nu_{\sigma}}\sqrt{ \frac{2}{\nu} }
                     \left[S_1^{\nu_{\sigma}}(|z|) + \nu_{\sigma} S_2^{\nu_{\sigma}}(0) - S_1^{\nu_{\sigma}}(0)  \right] \; .
 \label{vEE}                     
\end{equation}
Also,
\begin{align}
u_{\text{x}}^{\sigma}(z)= & \frac{-1}{r_s \nu_{\sigma}}\sqrt{ \frac{2}{\nu} }\,S_{1}^{\nu_{\sigma}}(|z|),
\label{orbital_potential}
\end{align}

\begin{align}
\bar{u}_{\text{x}}^{\sigma}=u_{\text{x}}^{\sigma}(0)= & \frac{-1}{r_s \nu_{\sigma}}\sqrt{ \frac{2}{\nu} }\,S_{1}^{\nu_{\sigma}}(0),
\end{align}

\begin{align}
\eta_{\text{x}}^{\sigma}= & -\frac{1}{r_{s}}\sqrt{\frac{2}{\nu}}\,S_{2}^{\nu_{\sigma}}(0).
\end{align}
In the last expressions we have  used that  $l_B={r_s}\sqrt{{\nu}/{2}}$, with $r_s$ being the 2D dimensionless parameter 
that characterizes the electronic density $N/A^*=(\pi r_s^2)^{-1}$. 
In the strict-2D limit the eigenvalues $\gamma_{1}^{\sigma}$ becomes the reference energy. 
We can also suppose that the LL broadening $\Gamma$ is smaller than the Zeeman splitting $\Delta E_z$ 
between spin-up and spin-down LL's, and that 
$\Delta E_z<\hbar\omega_{c}$ (that is $\hbar \omega_c>\Delta E_z>\Gamma \gg k_B T$). 
Then the LL's will be filled in the sequential order 
$(n=0,\uparrow)\rightarrow (n=0,\downarrow) \rightarrow (n=1,\uparrow) \rightarrow (n=1,\downarrow)...$, as we will 
consider in the following. 
We can write $S_{1}^{\nu_{\sigma}}(0)$ and $S_{2}^{\nu_{\sigma}}(0)$
in a simpler and more intuitive form:

\begin{align}
 S_{1}^{\nu_{\sigma}}(0) = & \sum_{n_{i},n_{j}=0}^{[\nu_{\sigma}]-1}I_{n_{i}}^{n_{j}}(0)+2p_{\sigma}\sum_{n_{i}=0}^{[\nu_{\sigma}]-1}I_{n_{i}}^{[\nu_{\sigma}]}(0)\nonumber \\
 & +p_{\sigma}^{2}I_{[\nu_{\sigma}]}^{[\nu_{\sigma}]}(0) \; , \nonumber \\
 \equiv & \; I_{1}([\nu_{\sigma}])+2p_{\sigma}I_{2}([\nu_{\sigma}])+p_{\sigma}^{2}I_{3}([\nu_{\sigma}])
 \label{eq:S1-simple} \; ,
\end{align}
and 

\begin{align}
 S_{2}^{\nu_{\sigma}}(0) = & 
 \sum_{n_{i}=0}^{[\nu_{\sigma}]-1}I_{n_{i}}^{[\nu_{\sigma}]}(0)+p_{\sigma}I_{[\nu_{\sigma}]}^{[\nu_{\sigma}]}(0)\nonumber \; , \\
 \equiv & \; I_{2}([\nu_{\sigma}])+p_{\sigma}I_{3}([\nu_{\sigma}]) \; .\label{eq:S2-simple}
\end{align}
Here 
\begin{eqnarray}
 I_{1}(n+1) = \int_{0}^{\infty}e^{-x^{2}/2}\left[L_{n}^{1}\left(\frac{x^{2}}{2}\right)\right]^{2}dx \; , \nonumber \\  
  = \frac{\sqrt{\pi}\;(2)_{n}(3/2)_{n}}{\sqrt{2}\;(n!)^2} 
 {}_{3}F _{2} (-n,\frac{1}{2},-\frac{1}{2};2,-n-\frac{1}{2};1) \; .
\label{eq:I1}
\end{eqnarray}
Also,
\begin{eqnarray}
 I_{2}(n) =  \int_{0}^{\infty}e^{-x^{2}/2}L_{n-1}^{1}\left(\frac{x^{2}}{2}\right)L_{n}^{0}\left(\frac{x^{2}}{2}\right)dx\;,
 \nonumber \\
 = \frac{\sqrt{\pi}\;(1)_{n}(3/2)_{n-1}}{\sqrt{2}\;n!(n-1)!} 
 {}_{3}F _{2} (-n,\frac{1}{2},-\frac{1}{2};1,-n+\frac{1}{2};1) \; , 
 \label{eq:I2}
\end{eqnarray}
and
\begin{eqnarray}
 I_{3}(n) = \int_{0}^{\infty}e^{-x^{2}/2}\left[L_{n}^{0}\left(\frac{x^{2}}{2}\right)\right]^{2}dx\;, \nonumber \\
 = \frac{\sqrt{\pi}\;(1)_{n}(1/2)_{n}}{\sqrt{2}\;(n!)^2} 
 {}_{3}F _{2} (-n,\frac{1}{2},\frac{1}{2};1,-n+\frac{1}{2};1) \; .
\label{eq:I3}
\end{eqnarray}
In passing from the first to the second line in Eqs. (\ref{eq:I1}), (\ref{eq:I2}), and (\ref{eq:I3}) we have used the 
result given in Ref. \cite{PBM92}. The $(a)_n$ are the Pochhammer's symbols  \cite{AS}, and then ${}_3F_2(a,b,c;d,e;z)$ is the 
(3,2) generalized hypergeometric function  \cite{AS} evaluated at $z=1$, for different values of the parameters $a,b,c,d,e$.
These explicit expressions for the quantities $I_1(n)$, $I_2(n)$, and $I_3(n)$ are very useful for the analytical
analysis of the zero-field limit, were $n \gg 1$.
Besides, in the last expressions we have used the identities 

\begin{align}
 \sum_{n_{i},n_{j}=0}^{[\nu_{\sigma}]-1}I_{n_{i}}^{n_{j}}(0)\text{=} & 
 \int_{0}^{\infty}e^{-x^{2}/2}\left[L_{[\nu_{\sigma}]-1}^{1}\left(\frac{x^{2}}{2}\right)\right]^{2}dx,
\end{align}

\begin{align}
 \sum_{n_{i}=0}^{[\nu_{\sigma}]-1}I_{n_{i}}^{[\nu_{\sigma}]}(0)= 
 & \int_{0}^{\infty}e^{-x^{2}/2}L_{[\nu_{\sigma}]-1}^{1}
 \left(\frac{x^{2}}{2}\right)L_{[\nu_{\sigma}]}^{0}\left(\frac{x^{2}}{2}\right)dx.
\end{align}

In the expression (\ref{eq:S1-simple}) for $S_{1}^{\nu_{\sigma}}(0)$
we can identify the first term as corresponding to the exchange energy between electrons
in filled LL's, the second term as the exchange energy
between electrons in filled LL's and in the (last) partially
filled LL, and the last term as representing the exchange energy between
electrons in the (last) partially filled LL.

As an interesting remark, it should be noted that the EE potential {\em at} the location of the strict-2D electron
gas is given by 
\begin{equation}
 v_{\text{x}}^{\sigma}(z=0) = -\frac{1}{r_{s}}\sqrt{\frac{2}{\nu}} \; S_{2}^{\nu_{\sigma}}(0) = \eta_{\text{x}}^{\sigma}\;.
 \label{vx0}
\end{equation}
As we will see later, it has a non-trivial magnetic field dependence through the function $S_{2}^{\nu_{\sigma}}(0)$.

As another piece of useful information, the one-particle density matrix for a strict-2D electron gas in a perpendicular
magnetic field can be defined as 
\begin{eqnarray}
 \rho_1^{\sigma}(\bm{\rho},\bm{\rho}')& = & 
 \sum_{n=0}^{[\nu_{\sigma}]-1}\sum_{k=1}^{N_{\phi}}\psi_{n,k}(\bm{\rho})^*\psi_{n,k}(\bm{\rho}') \nonumber \\
 &+&
 \sum_{k}^{occ.}\psi_{[\nu_{\sigma}],k}(\bm{\rho})^*\psi_{[\nu_{\sigma}],k}(\bm{\rho}'),
 \label{eq:rho_def}
\end{eqnarray}
The first term gives the contribution from all fully occupied LL, while the second term represents the contribution
from the last LL, whose occupation may be fractional.
Regarding this last term, and considering that within a given LL all values of $k$ are equally probably, 
we replace the sum over $k$ by an average over all $N_c$ possible configurations $\{k\}$:
\begin{align}
 \sum_k^{occ.}f_k(\sigma) \Rightarrow \frac{1}{N_c} \sum_{\{k\}}^{N_c} \sum_{k \in \{k\}} f_k(\sigma)
 =p_\sigma \sum_{k=1}^{N_{\phi}} f_k(\sigma),
 \label{average}
\end{align}
where $p_\sigma$ is the same occupation factor of the highest LL as it was previously defined. 
Substituting the sum over $k$ in (\ref{eq:rho_def}) we have
\begin{align}
 \rho_1^{\sigma}(\bm{\rho},\bm{\rho}')= & 
 \sum_{n=0}^{[\nu_\sigma]-1}\sum_{k=1}^{N_\Phi}\psi_{n,k}(\bm{\rho})^*\psi_{n,k}(\bm{\rho}') \nonumber \\ 
 & +p_\sigma\sum_{k=1}^{N_\Phi}\psi_{[\nu_\sigma],k}(\bm{\rho})^*\psi_{[\nu_\sigma],k}^{\sigma}(\bm{\rho}').
\end{align} 
Summing over all $k$ and  $n$   we obtain 
\begin{align}
 \rho_1^{\sigma}(\bm{\rho},\bm{\rho}')=& \frac{N_{\phi}}{A^*}e^{i\frac{(x'+x)(y'-y)}{2l_{B}^{2}}}e^{-\frac{|\mathbf{r}-
 \mathbf{r}'|^{2}}{4l_{B}^{2}}}
 \left[ L_{[\nu_{\sigma}]-1}^{1}\left(\frac{|\bm{\rho}-\bm{\rho}'|^{2}}{2l_{B}^{2}}\right) \right. \nonumber \\ 
 & \left. + \; p_\sigma L_{[\nu_{\sigma}]}^{0}\left(\frac{|\bm{\rho}-\bm{\rho}'|^{2}}{2l_{B}^{2}}\right) \right]\; .
 \label{md}
\end{align}
We can see from this last equation that the electron density
$\rho_1^{\sigma}(\bm{\rho},\bm{\rho}) = (N_{\phi}/A^*)\nu_{\sigma} = n_{\sigma}$,
since $L_{[\nu_{\sigma}]-1}^1(0)=[\nu_{\sigma}]$ and $L_{[\nu_{\sigma}]}^0(0) = 1$.
Note that the electron density is constant, that is, the electron
gas with an applied magnetic field  is homogeneous. Nevertheless we will
see in the following that there exist differences is the exchange
energy and potential for the homogeneous electron gas with and without
applied magnetic field. Eq. (\ref{md}) generalizes an equivalent expression for the one-particle density-matrix 
given in Ref. \cite{GV05},
restricted to the case of $p_{\sigma}=0$.

\subsection{The LSDA for the strict-2D electron gas at zero and finite magnetic field}

The exchange energy per particle for the arbitrarily spin-polarized 2D homogeneous electron gas at zero magnetic field is  \cite{GV05},
\begin{equation}
 e_{\text{x}}^{\text{LSDA}}(r_s,p) =  -\frac{2\sqrt{2}}{3 \pi r_{s}}\left[ (1+p)^{3/2}+(1-p)^{3/2}\right] \; , 
 \label{eq:ex-LSDA}
\end{equation}
where $p=(n_{\uparrow}-n_{\downarrow})/(n_{\uparrow}+n_{\downarrow})$ is the fractional spin polarization. For inhomogeneous
2D systems in the presence
of a perpendicular magnetic field $B$, and in the Local Spin Density Approximation (LSDA), the same expression is
assumed to be valid, but for the magnetic-field and position-dependent densities $n_{\uparrow}(\bm{\rho})$ and 
$n_{\downarrow}(\bm{\rho})$.
It is one of the main goals of the present work to analyze the validity of the LSDA, 
for the case of homogeneous 2D systems in the IQHE regime, by comparison with our EE results.

The LSDA exchange potential for a strict-2D homogeneous electron gas at zero
magnetic field can be obtained from the exchange energy per particle
using that 
\begin{equation}
 v_{\text{x}}^{\sigma,\text{LSDA}} \equiv \partial(n e_{\text x}^{\text{LSDA}})/\partial n_{\sigma} =
 -\frac{2\sqrt{2}}{\pi r_{s}}\sqrt{1\pm p} \; ,
 \label{vLSDA}
\end{equation}
where we should take the $+$ sign for the spin-up component and the $-$ sign for
the spin-down component. By assuming again that $r_s$ and $p$ are position and magnetic-field dependent quantities
through their density dependence, Eq. (\ref{vLSDA}) becomes the strict-2D LSDA for the exchange potential.
Eqs. (\ref{eq:Ex_2D_frac_fill_factor}) and (\ref{vEE}) are not obviously related to their LSDA counterparts given by 
Eqs. (\ref{eq:ex-LSDA}) and (\ref{vLSDA}). We will shown later analytically and numerically that the LSDA expressions are
the leading contributions of the corresponding EE rigorous expressions, in the limit of small magnetic field.

\section{Results and discussions} 
\label{secc:3}

\subsection{Comparison between exact exchange and LSDA at finite magnetic field}

\begin{figure}[h]
 \includegraphics[width=8.8cm]{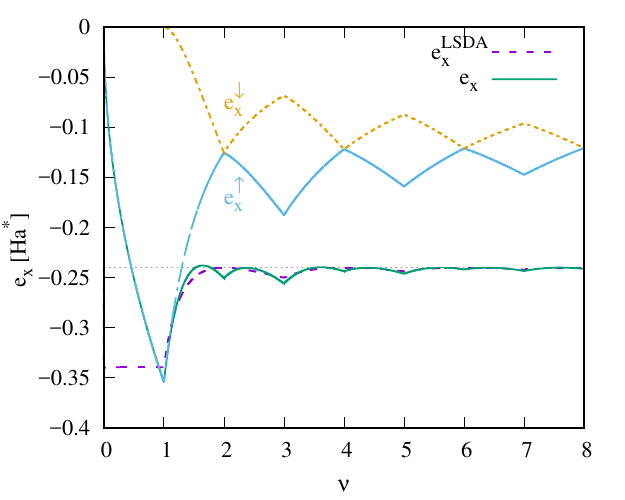}\caption{\label{fig:exvsnu}
 Exact-exchange energy per particle $e_{\text{x}}(r_s,\nu)$ and its local spin density approximation
 $e_{\text{x}}^{\text{LSDA}}(r_s,p(\nu))$ vs $\nu$, for $r_s = 2.5$. The horizontal line at $-4\sqrt{2}/(3 \pi r_s) \simeq -\;0.24$
 corresponds to the common $\nu \gg 1$ limit of $e_{\text{x}}(r_s,\nu)$ and $e_{\text{x}}^{\text{LSDA}}(r_s,p(\nu))$.
 The spin-discriminated contributions $e_{\text{x}}^{\uparrow}(r_s,\nu)$ and $e_{\text{x}}^{\downarrow}(r_s,\nu)$
 are also displayed.}
\end{figure}

Fig.~\ref{fig:exvsnu} shows the EE energy per particle vs $\nu$, for $r_s = 2.5$~  \cite{td}.
The LSDA result is also shown for comparison, using expression (\ref{eq:ex-LSDA}) and
making the replacement $ p \rightarrow p(\nu) = [\nu_{\uparrow}(\nu)-\nu_{\downarrow}(\nu)]/\nu$~  \cite{note1}.
The differences between both results increases when the magnetic field is increased
(small $\nu$ limit). The exact-exchange energy displays derivative discontinuities
at every integer filling factor $\nu$, while this kind of discontinuity
is present in the LSDA energy only at odd integer values of $\nu$. In the
LSDA the derivative of the exchange energy can be written as 
${d e_{\text{x}}^{\text{LSDA}}}/{d\nu}={d e_{\text{x}}^{\text{LSDA}}}/{dp}\times{dp(\nu)}/{d\nu}$,
and the discontinuities enter through ${dp}/{d\nu}$ that is discontinuous
at every integer value of $\nu$. However, at even $\nu$ values ${d e_{\text{x}}^{\text{LSDA}}}/{dp}=0$,
the effect of ${dp(\nu)}/{d\nu}$ is lost, and the derivative is
continuous at this points. The difference between the LSDA and EE is rooted in the fact that the LSDA
is a zero magnetic-field approximation, reflected in the parity property that 
$e_{\text{x}}^{\text{LSDA}}(r_s,p) = e_{\text{x}}^{\text{LSDA}}(r_s,-p)$.
This means that their expansion in the $p \rightarrow 0$ limit only involves even powers of $p$. This leads to
the property ${d e_{\text{x}}^{\text{LSDA}}}/{dp}=0$ for even $\nu$. On the other side, 
$e_{\text x}(r_s,\nu)$ fully includes the effect of the magnetic field.
Other interesting difference between these two approaches to the exchange
energy is that the EE result presents local minima at every integer $\nu$
values, while the LSDA exchange energy has local minima only at odd $\nu$ and local
maxima at even $\nu$. In the LSDA approach the exchange energy only
depends (at constant density) on the polarization $p$, and the polarization
presents local minima at even $\nu$ (loosing exchange energy) and
local maxima at odd $\nu$'s (gaining exchange energy). 
The behavior of EE energy is however more complicated: besides $p$, it also depends
of the occupation factor of the last occupied LL. 
The spin-discriminated EE energies are also displayed in Fig.~\ref{fig:exvsnu}.
$e_{\text{x}}^{\uparrow}(r_s,\nu)$ decreases when $[\nu] < \nu < [\nu]+1$, with $[\nu]=$ even, while it increases
for $[\nu] < \nu < [\nu]+1$ when $[\nu]=$ odd. $e_{\text{x}}^{\downarrow}(r_s,\nu)$ shows the opposite behavior.
Both $e_{\text{x}}^{\uparrow}(r_s,\nu)$ and $e_{\text{x}}^{\downarrow}(r_s,\nu)$ have large derivative
discontinuities of opposite signs at every integer $\nu$, that nearly compensates to give the much weaker,
but still finite
slope discontinuities of $e_{\text{x}}(r_s,\nu)$.

\begin{figure}[h]
 \includegraphics[width=8.8cm]{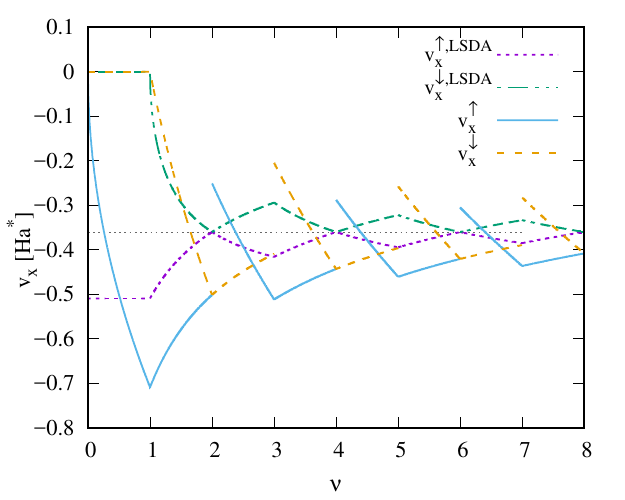}
 \caption{\label{fig:vxvsnu}Exact-exchange potential $v_{\text x}^{\sigma}(0)$ and its LSDA counterpart 
 $v_{\text x}^{\sigma, {\text LSDA}}$ at $z=0$ vs $\nu$, for $r_s=2.5$.
 The horizontal line at $-2\sqrt{2}/(\pi r_s) \simeq -\;0.36$
 corresponds to the common $\nu \gg 1$ limit of $v_{\text x}^{\sigma}(0)$ and $v_{\text x}^{\sigma,\text{LSDA}}$.}
\end{figure}

We display in Fig.~\ref{fig:vxvsnu} the EE and LSDA exchange potentials for both spin components.
$v_{\text x}^{\uparrow}(0)$ has sharp discontinuities at every even $\nu$, while
$v_{\text x}^{\downarrow}(0)$ has the discontinuities located at odd values of $\nu$,
excepting $\nu=1$. All these abrupt jumps are related to the filling of a new LL,
as we will discuss below.
On the other side, both components of the LSDA exchange potential are continuous, and only exhibits derivative
discontinuities at every integer filling factor. $v_{\text{x}}^{\uparrow,\text{LSDA}}$ is constant for $0<\nu<1$, since for this
strong field regime all electrons are fully spin-polarized in the ground LL, with $p=1$.
The derivative discontinuities of $v_{\text{x}}^{\sigma,\text{LSDA}}$ are just a consequence of its
dependence on $p(\nu)$, that also has a discontinuous derivative at every integer $\nu$.
As has been already pointed out, the discontinuities in $v_{\text x}^{\sigma}(0)$ at integer $\nu$'s are due to the 
non-trivial behavior with magnetic field of the function $S_2^{\nu_{\sigma}}(0)$ in Eq. (\ref{vEE}) or equivalently
the function $\eta_{\text x}^{\sigma}$ in Eq. (\ref{vx0}). 
Proceeding from this last equation, the abrupt jump in the EE exchange potential at every integer $\nu$ may be
written exactly as
\begin{align}
 \Delta v_{\text x}^{\sigma}= & \frac{-1}{r_{s}}\sqrt{\frac{2}{\nu}}
 \left[S_{2}^{[\nu_{\sigma}]^{+}}(0)-S_{2}^{[\nu_{\sigma}]^{-}}(0)\right]\nonumber \; ,\\
 = & \frac{-1}{r_{s}}\sqrt{\frac{2}{\nu}}
 \left[I_{2}([\nu_{\sigma}])-I_{2}([\nu_{\sigma}]-1)-I_{3}([\nu_{\sigma}]-1)\right] \; ,\nonumber \\
 = & \frac{1}{r_{s}}\sqrt{\frac{2}{\nu}}
 \int_{0}^{\infty}e^{-X}L_{[\nu_{\sigma}]-1}^{1}\left(X\right) \nonumber \\
 & \times \left[L_{[\nu_{\sigma}]-1}^{0}\left(X \right)-L_{[\nu_{\sigma}]}^{0}
  \left(X \right)\right] \;dx \; ,
\end{align}
with $X=x^2/2$.
We have checked numerically that the last integral is finite and positive, for finite $[\nu_{\sigma}]$.
On the other side, for $[\nu_{\sigma}] \gg 1$, the difference between the two generalized Laguerre polynomials
inside the integral is increasingly small, and then the discontinuity in the EE also vanishes asymptotically in the 
low-field limit.
Note however the quite different behavior of $e_{\text x}(r_s,\nu)$ and of $v_{\text x}^{\sigma}(0)$ in the large
$\nu$ limit: while for $\nu \sim 10$ $e_{\text x}(r_s,\nu)$ and $e_{\text x}^{\text{LSDA}}(r_s,p(\nu))$ are
essentially indistinguishable on the drawing scale in Fig.~\ref{fig:exvsnu}, the difference between 
$v_{\text x}^{\sigma}(0)$ and $v_{\text x}^{\sigma,\text{LSDA}}$ are still clearly discernible in Fig.~\ref{fig:vxvsnu}
for large $\nu$. This point will be discussed in more detail in the following section. It should be also emphasized
that the jump given by $\Delta v_{\text x}^{\sigma}$ above applies also to the full exchange potential
$v_{\text x}^{\sigma}(z)$, and not only to its strict-2D contribution $v_{\text x}^{\sigma}(0)$ discussed above.
This issue is further discussed in the discussion surrounding Fig.~\ref{fig:VxvsZ}.

\subsection{Zero magnetic-field limit}

From the previous results we can see that by decreasing the magnetic field
(and then increasing $\nu$) the EE results somehow become similar to the LSDA
results. Now we will shown  that we can obtain the LSDA results analytically
as a zero-field limit of our EE expressions. This can be considered as a critical test on the consistency
of the present formalism. In the first place, the EE energy in expression (\ref{eq:Ex_2D_frac_fill_factor})
should reduce to expression (\ref{eq:ex-LSDA}). For proving that, we can write
the EE energy per particle in the form, 

\begin{align}
 e_{\text x}(r_s,\nu) =  -\frac{2\sqrt{2}}{3 \pi r_{s}}
 \left\{ [1+p(\nu)]^{3/2}\bar{S}_1^{\nu_{\uparrow}}+[1-p(\nu)]^{3/2}\bar{S}_1^{\nu_{\downarrow}}\right \} \; ,
\end{align}
with $\bar{S}_1^{\nu_{\sigma}} = (3\pi/8\sqrt{2})S_1^{\nu_{\sigma}}(0)/\nu_{\sigma}^{3/2}$. This re-writing is
motivated by the fact that now the difference between the EE and LSDA exchange energies is just related to how far
are the factors $\bar{S}_1^{\nu_{\sigma}}$ from unity. Using that (see Appendix \ref{appendix:limits}, Eq. (\ref{eq:S1_A1}))
\begin{align}
 \lim_{\nu_{\sigma}\rightarrow\infty} \bar{S}_1^{\nu_{\sigma}}  
 \rightarrow & 1 \; ,
 \label{S1limit}
\end{align}
we have 
\begin{equation}
 \lim_{\nu\rightarrow\infty} e_{\text x}(r_s,\nu) = 
 e_{\text x}^{ {\text {LSDA}}}(r_s,p) \; ,
 \label{elimit}
\end{equation}
where we have used that $p(\nu) = [\nu_{\uparrow}(\nu)-\nu_{\downarrow}(\nu)]/\nu$.
This last result is the well-know exchange energy of the strict-2D electron gas
at zero magnetic-field presented above. 

On the other hand the limit $\nu \gg 1$ of the EE potential at $z=0$, that is, the potential
at the electron gas should coincide also with the respective strict-2D
zero-field LSDA result. For the EE potential we can work in a similar way as with the energy,
doing a re-writing of Eq. (\ref{vx0})
\begin{align}
 v_{\text{x}}^{\sigma}(z=0) = -\frac{2\sqrt{2}}{\pi r_{s}} \left[1 \pm p(\nu) \right]^{1/2} \bar{S}_2^{\nu_{\sigma}} \; ,
 \label{vEE*}
\end{align}
with $\bar{S}_2^{\nu_{\sigma}} = (\pi/2\sqrt{2})S_2^{\nu_{\sigma}}(0)/\nu_{\sigma}^{1/2}$.
Using now that (see Appendix \ref{appendix:limits}, Eq. (\ref{eq:S2_A2}))
\begin{align}
\lim_{\nu_{\sigma}\rightarrow\infty} \bar{S}_{2}^{\nu_{\sigma}} \rightarrow 1 &  \; ,
\end{align}
we obtain 
\begin{align}
 \lim_{\nu\rightarrow\infty} v_{\text x}^{\sigma}(z=0)= v_{\text x}^{\sigma,{\text {LSDA}}} \; ,
 \label{vlimit}
\end{align}
the same expression as in the zero-field case. These two quantities $\bar{S}_1^{\nu_{\sigma}}$ and 
$\bar{S}_2^{\nu_{\sigma}}$ may be considered as a sort of finite-$\nu$ ``correction'' factors, whose inclusion brings the LSDA results
towards the EE expressions.

In Fig.~3 we shown how these two ``correction'' factors approach the LSDA limit as $\nu$ increases,
for the spin-up case. While it is seen
there that $\bar{S}_1^{\nu_{\uparrow}}$ approach the large $\nu$ limit of Eq. (\ref{S1limit}) quite rapidly, the approach
of $\bar{S}_2^{\nu_{\uparrow}}$ to the LSDA limit is much slower, that explains the persistence of sizeable
discontinuities in $v_{\text x}^{\uparrow}(0)$ in Fig.~2, even for large values of $\nu$.

The abrupt jump expression for $\Delta v_{\text x}^{\sigma}$ may be also analyzed by using the asymptotic
expansions for $I_2(n)$ and $I_3(n)$ given in the Appendix. Since $I_2([\nu_{\sigma}]) \sim ([\nu_{\sigma}])^{1/2}$
and $I_3([\nu_{\sigma}]) \sim \ln([\nu_{\sigma}])/([\nu_{\sigma}])^{1/2}$ for large $\nu_{\sigma}$, in the same
low-field limit $\Delta v_{\text x}^{\sigma}$ vanishes asymptotically as 
$\ln([\nu_{\sigma}])/[\nu_{\sigma}]$. This is consistent with the fact that
the LSDA exchange potential, that corresponds to the $\nu_{\sigma} \rightarrow \infty$ limit 
has no discontinuities at integer filling factors.

\begin{figure}[h]
\includegraphics[width=8.8cm]{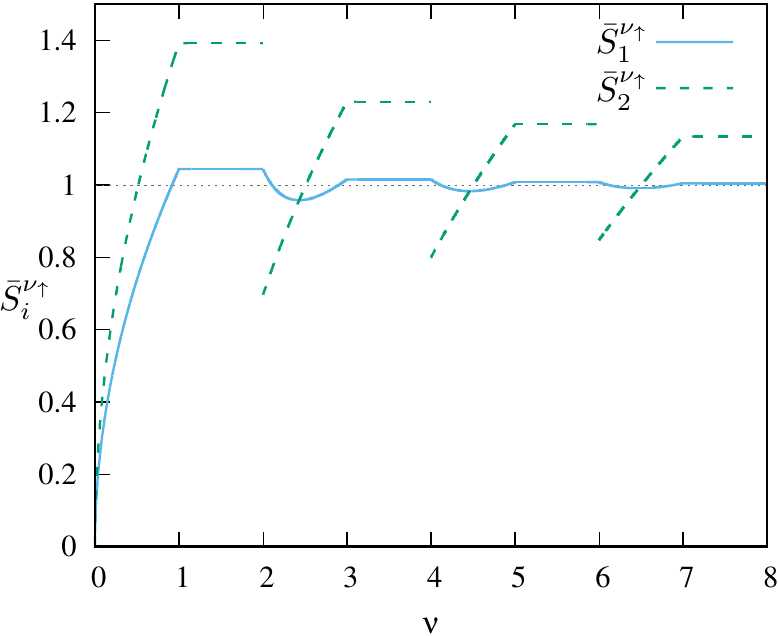}
 \caption{\label{fig:S1S2vsnu} $\bar{S}_1^{\nu_{\uparrow}}$ and $\bar{S}_2^{\nu_{\uparrow}}$ as function of the filling
 factor $\nu$. The LSDA limit at $\nu \gg 1$ is represented by the horizontal line at unity, and 
 it is seen as they approach this limit in quite different ways. $\bar{S}_2^{\nu_{\uparrow}}$ has an abrupt jump
 at every even $\nu$.}
\end{figure}

It is important to remark the importance of the inclusion of $\eta_{\text{x}}^{\sigma}$
term, which comes from the implicit derivative of the exchange energy
with respect to the density. This term is the origin of derivative
discontinuity and without it we cannot obtain the correct zero magnetic-field limit.

\subsection{On the z-dependence of $v_{\text x}^{\sigma}(z)$}

Up to this point we have only considered the EE potential at the electron gas coordinate $z=0$.
But as we have already discussed, the present procedure has the advantage that it provides quite naturally
also its $z$-dependence, as follows.

Fig.~4 displays the spin-up EE potential as defined in Eq. (\ref{vEE}), as a function of $z$ and for several
values of $\nu$, either approaching an even integer filling factor from above ($\nu^+$), or from below ($\nu^-$).
Several interesting features of Fig.~4 are worth of be commented: i) it gives a more complete perspective about
how the EE potential discontinuity present at every integer $\nu$ at $z=0$ and displayed in Fig.~2 evolves with
the distance $z$; ii) while the EE potentials corresponding to $\nu^+$ approach the zero-field limit from
above for increasing values of $\nu$, the ones corresponding to $\nu^-$ approach the same limit from below;
and iii) in the asymptotic limit $|z|/l_B \gg 1$, the EE potential approach a finite non-negative value,
that depends on the density and the magnetic field.

\begin{figure}[h]
\includegraphics[width=8.8cm]{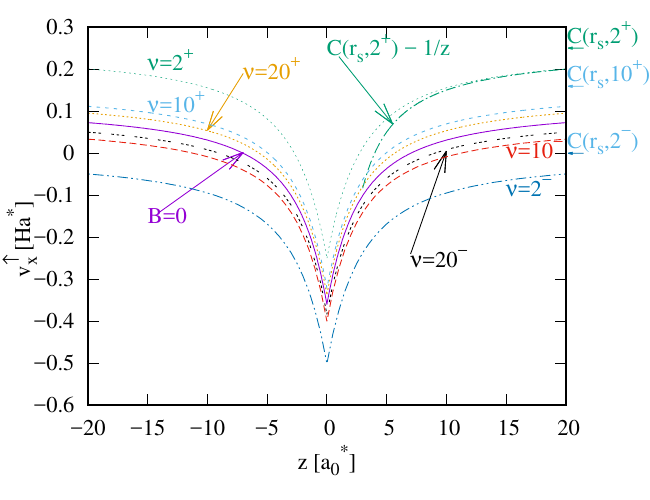}
 \caption{\label{fig:VxvsZ}Exact-exchange potential vs $z$ for different filling factors. 
 The EE potential for the spin-unpolarized 2D electron gas at zero magnetic field is also shown for comparison. 
 The result for $\nu=2^+$ is compared with the expected asymptotic behavior for large $z$.
 The asymptotic limits for same of the potentials are given on the right vertical axis, and marked with arrows.}
\end{figure}

This last point can be further elaborated analytically: the large $z$-limit of $v_{\text x}^{\sigma}(z)$ is
given by the large $z$-limit of $S_1^{\nu_{\sigma}}(|z|)$ in Eq. (\ref{vEE}). Now, according Eq. (\ref{DefS1}),
what matters for this limit is how the functions $I_n^m(t)$ behave for large values of their argument.
Inspection of Eq. (\ref{I_nm}) reveals that for $|z-z'|/l_B \gg 1$, only the limit of the integrand for small
values of $x$ contributes to the integral. Considering that $L_{n_<}^{n_>-n_<}(0)=n_>!/n_<!(n_>-n_<)!$, we have then 
\begin{eqnarray}
 I_n^m\left(\frac{|z|}{l_B} \gg 1\right) &\rightarrow& 
 \frac{n_>!}{n_<![(n_>-n_<)!]^2} \nonumber \\
 &\times& \int_0^\infty \left(\frac{x^2}{2}\right)^{n_>-n_<}e^{-x|z|/l_B}dx \nonumber \\
 = \frac{n_>![2(n_>-n_<)]!}{n_<![(n_>-n_<)!]^2} &\times& \frac{1}{2^{(n_>-n_<)}} 
 \left(\frac{l_B}{|z|}\right)^{1+2(n_>-n_<)}  \; . \nonumber \\
\end{eqnarray}
From this last expression is clear that the leading contribution to $I_n^m(|z|/l_B \gg 1)$
corresponds to $n_> = n_<$, and then to $n=m$: 
$I_n^m(|z|/l_B \gg 1) \rightarrow I_n^n(|z|/l_B \gg 1) = l_B/|z|$.
And since $\sum_n^{[\nu_\sigma]-1} I_n^n(|z|/l_B \gg 1)=[\nu_\sigma]$ we obtain 
\begin{equation}
 v_{\text{x}}^{\sigma}(|z|/l_B \gg 1) \rightarrow C(r_s,\nu) - \frac{[\nu_\sigma]+p_\sigma^2}{z\nu_\sigma} \; , 
\end{equation}
where 
\begin{equation}
C(r_s,\nu)=\frac{1}{r_s \nu_{\sigma}}\sqrt{ \frac{2}{\nu} }
                     \left[S_1^{\nu_{\sigma}}(0)-\nu_{\sigma} S_2^{\nu_{\sigma}}(0)   \right] \; .
\end{equation}
We have checked numerically that the term $C(r_s,\nu)$ is always non-negative; besides, it is equal to zero only 
when $0<\nu_\sigma<1$  \cite{constant}. 
Interestingly we obtain the universal $1/z$ asymptotic behavior  \cite{RHP15} only in the limits of  
$p_\sigma \rightarrow (0,1)$  \cite{nuinteger}. 
In Fig.~\ref{fig:VxvsZ} we compare this asymptotic behavior with the EE potential for $\nu=2^+$,
and it is seen how the asymptotic limit is reached for $|z|/l_B \sim 10$.  

For the sake of completeness, we provide here also the spin-polarized EE potential at zero-magnetic field, which is given by  \cite{HPR97} 
\begin{align}
 v_{\text x}^{\sigma,B=0}(z)= & -\frac{1}{|z|}\left(1+\frac{L_{1}(2|z|^*)-I_{1}(2|z|^*)}
 {|z|^*}\right)\nonumber \\
 & + \frac{2\sqrt{2}}{3\pi r_s} \sqrt{1 \pm p} \; ,
 \label{vzf}
\end{align}
with $L_1(x)$ and $I_1(x)$ being the Struve and the modified Bessel functions, respectively,
and $|z|^*=\sqrt{2(1 \pm p)} |z| / r_s$.
This expression has been obtained from the quasi-2D EE integro-differential equation at zero magnetic field,
after imposing the same one-subband constraint and strict-2D limit as in the present contribution. Its derivation will be given
elsewhere. While in principle it is feasible to obtain it by taking the zero-field limit of Eq. (\ref{vEE}), the lack of
the simplifying identities in Eqs. (\ref{eq:I1}), (\ref{eq:I2}), and (\ref{eq:I3}) makes all the calculations much
more involved. Using that the limit of $1 + (L_1(2x) - I_1(2x))/x$ for $x \rightarrow 0$ is given by $8x/3\pi$, by
evaluating the zero-field EE potential at $z=0$ one obtains
\begin{equation}
 v_{\text x}^{\sigma,B=0}(z=0) = - \frac{2\sqrt{2}}{\pi r_s} \sqrt{1 \pm p} \; ,
\end{equation}
that coincides with the zero-field limit of Eq. (\ref{vlimit}), as it should be. It is important to note that this
internal consistency is only possible if the constant term $\frac{2\sqrt{2}}{3\pi r_s} \sqrt{1 \pm p}$ is present
in Eq. (\ref{vzf}).

The $z$-dependence of the EE potential can be useful for the study of coupled 2D electron
gases like bilayers  \cite{HMD00} and trilayers  \cite{HM96,MP16,WMGRBP09,MPB16}, in which this dependence appears as an inter-layer
exchange term. 

\section{Conclusions}
Starting from a general exact-exchange quasi-2D formalism describing an electron gas confined in a semiconductor
quantum well in the regime of the integer quantum Hall effect, we have obtained its strict-2D projection. The 
corresponding strict-2D calculations are much simpler than the ones associated with the quasi-2D case, since in the 
latter situation the self-consistent Kohn-Sham orbitals describing the subband physics must be obtained numerically.
Instead, in our strict-2D evaluation, these Kohn-Sham orbitals are replaced by Dirac delta-functions, that constraints
the electron dynamics to a single plane.
As the filling of the emerging Landau
levels proceeds, two main features results: {\it i)} the EE energy minimizes with a discontinuous derivative at every
integer filling factor $\nu$; and {\it ii)} the EE potential display sharp discontinuities at every integer $\nu$.
On the other side, the standard strict-2D LSDA displays derivative discontinuities in the exchange energy only at odd
filling factors, and has no discontinuities in the corresponding exchange potential. It should be emphasized that
these important differences between LSDA and EE at finite magnetic fields are present even when both are based in the 
same density, that remains homogeneous at finite magnetic field. More to the point, the differences are due to the 
fact that the functional form of the EE energy fully includes the effect of the magnetic field through the Landau orbitals,
while the LSDA is a zero-field approximation.

The present work suggest, however, in a very natural way how to go beyond the strict-2D LSDA as applied in the strong
magnetic field regime of inhomogeneous two-dimensional electron systems: replace $v_{\text x}^{\sigma,{\text {LSDA}}}$
of Eq. (\ref{vLSDA}) by $v_x^{\sigma}(0)$ of Eq. (\ref{vEE*}). In the standard LSDA, the zero-field expression for the homogeneous
2D electron gas, $v_{\text x}^{\sigma,{\text {LSDA}}}$ is applied to the finite magnetic field and inhomogeneous
case by doing the local-density-approximation, that amounts to the replacements 
$r_s \rightarrow r_s(\bm{\rho}) = 1/\sqrt{\pi n(\bm{\rho})}$, and
$p \rightarrow p(\bm{\rho}) = (n_{\uparrow}(\bm{\rho})-n_{\downarrow}(\bm{\rho}))/n(\bm{\rho})$.
Under these assumptions, $v_{\text x}^{\sigma,{\text {LSDA}}}$ becomes a position-dependent potential
$v_{\text x}^{\sigma,{\text {LSDA}}} \rightarrow v_{\text x}^{\sigma,{\text {LSDA}}}(\bm{\rho})$.
We suggest here that a more founded procedure will be to do the same local-density-approximation, but for our
EE potential: $v_{\text x}^{\sigma}(0) \rightarrow v_{\text x}^{\sigma}(\bm{\rho},0)$. The crucial difference
between $v_{\text x}^{\sigma,{\text {LSDA}}}(\bm{\rho})$ and $v_{\text x}^{\sigma}(\bm{\rho},0)$ is that while
in the former the effect of the magnetic field enters {\it indirectly} through the field-induced changes in the spin-polarization,
in $v_{\text x}^{\sigma}(\bm{\rho},0)$ the impact of the field is direct, since the {\it functional form} of the EE
already fully contains the magnetic field. In other words, while $v_{\text x}^{\sigma}(\bm{\rho},0)$ is exact in the 
case of the uniform 2D gas at arbitrary magnetic field, $v_{\text x}^{\sigma,{\text {LSDA}}}(\bm{\rho})$ only is
exact for the case of the homogeneous 2D gas at zero magnetic-field.
\label{secc:conclu}

\section{Acknowledgements}
DM and CRP acknowledge the financial support of ANPCyT under grant PICT 2016-1087, and CONICET under
grant PIP 2014-2016 00402.

\appendix

\section{Zero field limits}
\label{appendix:limits}
\setcounter{equation}{0}
\numberwithin{equation}{section}
The analysis of the zero magnetic-field limit amounts to study the large $n$-limit of the functions
$I_1(n)$, $I_2(n)$, and $I_3(n)$, defined in Eqs. (\ref{eq:I1}), (\ref{eq:I2}), and (\ref{eq:I3}), respectively.

Starting with $I_1(n)$, we have that in the large-$n$ limit
\begin{eqnarray}
 {}_3F_2\left(-n,\frac{1}{2},-\frac{1}{2};2,-n-\frac{1}{2};1 \right) = \sum_{k=0}^{\infty}
 \frac{(-n)_k(\frac{1}{2})_k(-\frac{1}{2})_k}{k!(2)_k(-n-\frac{1}{2})_k} \; , \nonumber \\
 \simeq \sum_{k=0}^{\infty}
 \frac{(\frac{1}{2})_k(-\frac{1}{2})_k}{(2)_k}\frac{1}{k!} = {}_2F_1\left(\frac{1}{2},-\frac{1}{2};2;1 \right) \; . 
 \nonumber \\
\end{eqnarray}
Using now the relation  \cite{AS}
\begin{equation}
 {}_2F_1\left(a,b;c;1 \right) = \frac{\Gamma(c)\Gamma(c-a-b)}{\Gamma(c-a)\Gamma(c-b)} \; ,
 \label{2F1} 
\end{equation}
we obtain that ${}_2F_1\left(\frac{1}{2},-\frac{1}{2};2;1 \right)= \frac{8}{3\pi}$. Collecting all results so far,
\begin{equation}
 I_1(n+1) \simeq \frac{8}{3\sqrt{2 \pi}} \frac{(2)_n(3/2)_n}{(n!)^2} \; .
\end{equation}
In the last step, since the Pochhammer's symbols admit the asymptotic expansion
\begin{equation}
 (a)_n \simeq \frac{\sqrt{2\pi}}{\Gamma(a)}e^{-n}n^{a+n-1/2}\left[1+ \mathcal{O}(1/n) \right] \; ,
\end{equation}
and using the Stirling approximation $n! \simeq \sqrt{2\pi}n^{n+1/2}e^{-n}$, we obtain that
\begin{equation}
 I_1(n+1) \simeq \frac{8\sqrt{2}}{3\pi}n^{3/2} + \mathcal{O}\left({n^{1/2}}\right) \; .
 \label{eq:l1}
\end{equation}

Regarding the large-$n$ limit of $I_2(n)$, we are now concerned with the asymptotic limit of
\begin{eqnarray}
 {}_3F_2\left(-n,\frac{1}{2},-\frac{1}{2};1,\frac{1}{2}-n;1 \right) = \sum_{k=0}^{\infty}
 \frac{(-n)_k(\frac{1}{2})_k(-\frac{1}{2})_k}{k!(1)_k(\frac{1}{2}-n)_k} \; , \nonumber \\
 \simeq \sum_{k=0}^{\infty}
 \frac{(\frac{1}{2})_k(-\frac{1}{2})_k}{(1)_k}\frac{1}{k!} = {}_2F_1\left(\frac{1}{2},-\frac{1}{2};1;1 \right) =
 \frac{2}{\pi}\; . 
 \nonumber \\
\end{eqnarray}
In the last step we have used Eq. (\ref{2F1}) again. Returning to Eq. (\ref{eq:I2}),
\begin{equation}
 I_2(n) \simeq \sqrt{\frac{2}{\pi}} \frac{(1)_n(3/2)_{n-1}}{n!(n-1)!} \simeq
 \frac{2\sqrt{2}}{\pi} n^{1/2} + \mathcal{O}\left(\frac{1}{n^{1/2}}\right) \; .
 \label{eq:l2}
\end{equation}
Regarding the large-$n$ limit of $I_3(n)$, we have numerical evidence that $I_3(n) \sim \ln(n)/n^{1/2}$ for $n \gg 1$.
The scalings $I_1(n) \sim n^{3/2}$ and $I_2(n) \sim n^{1/2}$ are easy of understand from the corresponding
definitions in Eq. (\ref{eq:S1-simple}): for a given scaling of $I_3(n)$, $I_2(n)$ has about $n$ contributions
more, while $I_1(n)$ has a double sum with about $n^2$ terms. This is understable also from a physical point of
view: the contribution to the EE energy and potential of the last fractionally occupied LL should be negligible
when the number of fully occupied LL becomes very large.

Putting all results together,
\begin{align}
 \lim_{\nu_{\sigma}\rightarrow\infty} \bar{S}_{1}^{\nu_{\sigma}}
 = & \frac{3\pi}{8\sqrt{2}} \lim_{\nu_{\sigma}\rightarrow\infty}  
 \frac{I_{1}([\nu_{\sigma}])+2p_{\sigma}I_{2}([\nu_{\sigma}])+p_{\sigma}^{2}I_{3}([\nu_{\sigma}])}{\nu_{\sigma}^{3/2}} \nonumber \\
 & \rightarrow 1 \; . 
 \label{eq:S1_A1} 
\end{align}
In a similar way we can show that 
\begin{align}
 \lim_{\nu_{\sigma}\rightarrow\infty}\bar{S}_{2}^{\nu_{\sigma}} =
 & \frac{\pi}{2\sqrt{2}} 
 \lim_{\nu_{\sigma}\rightarrow\infty}\frac{(I_{2}([\nu_{\sigma}])+p_{\sigma}I_{3}([\nu_{\sigma}])}{\nu_{\sigma}^{1/2}} \nonumber \\
 \rightarrow & 1.
 \label{eq:S2_A2}
\end{align}
In principle, from the corrections to the leading term in Eqs. (\ref{eq:l1}) and (\ref{eq:l2}) 
is feasible to obtain the corresponding corrections to the leading (LSDA) limits for $e_{\text x}(r_s,\nu)$ and
$v_{\text x}^{\sigma}(z)$, as given by Eqs. (\ref{elimit}) and (\ref{vlimit}), respectively.
Since the final results are somehow involved, they will be given elsewhere.


\begin{thebibliography}{}

\bibitem{WvK11} J. Weis and K. von Klitzing, Philos. Trans. R.Soc. A \textbf{369}, 3954 (2011).

\bibitem{dSP97} S. das Sarma and A. Pinczuk, in {\it Perspectives in Quantum Hall Effects} (Wiley,New York,1997).

\bibitem{GV05}G. F. Giuliani and G. Vignale in
{\it Quantum Theory of the Electron Liquid}, Cambridge University Press,
Cambridge, (2005).

\bibitem{MFP17} D. Miravet, G. J. Ferreira and C. R. Proetto, Europhys.
Lett. \textbf{119},\textbf{ }57001 (2017).

\bibitem{PY89} R. G. Parr and W. Yang, in {\it Density Functional Theory of Atoms and Molecules} 
 (Oxford University Press, New York, 1989); R. M. Dreizler and E. K. U. Gross, in {\it Density Functional Theory} 
 (Springer, Berlin, 2000).
 
 \bibitem{G00} T. Grabo, T. Kreibich, S. Kurth, and E. K. U. Gross in {\it Strong Coulomb Interactions in Electronic Structure Calculations: Beyond the Local
 Density Approximation}, edited by V. I. Anisimov (Gordon and Breach, Amsterdam, 2000). 

\bibitem{HKPRG08} S. Pittalis, S. Kurth, N. Helbig, and E. K. U. Gross, Phys. Rev. A \textbf{74}, 065511 (2006); 
 S. Sharma, J. K. Dewhurst, C. Ambrosch-Draxl, S. Kurth, N. Helbig, S. Pittalis, S. Shallcross,
 L. Nordstr\"om, and E. K. U. Gross, Phys. Rev. Lett. \textbf{98}, 196405 (2007);
 N. Helbig, S. Kurth, S. Pittalis, E. R\"as\"anen, and E. K. U. Gross, Phys. Rev. B \textbf{77}, 245106-1 (2008).

\bibitem{BKMPLMM11} S. Becker, C. Karrasch, T. Mashoff, M. Pratzer, M. Liebmann, V. Meden, and M. Morgenstern,
 Phys. Rev. Lett. \textbf{106}, 156805 (2011).
 
\bibitem{B88} G. Bastard, in {\it Wave mechanics applied to Semiconductor Heterostructures} 
 (Les Editions de Physique, Les Ulis, 1988).

\bibitem{AS17} E. R\"as\"anen, H. Saarikoski, A. Harju, M. Ciorga, and A. S. Sacharjda, 
 Phys. Rev. B \textbf{77}, 041302(R) (2008).; M. C. Rogge, E. R\"as\"anen, and R. J. Haug, Phys. Rev. Lett.
 \textbf{105} 046802 (2010); H. Atci, U. Erkarslan, A. Siddiki, and E. R\"as\"anen, J. Phys.: Condens. Matter 
 \textbf{25} 155604 (2013); H. Atci and A. Siddiki, Phys. Rev. B \textbf{95}, 045132 (2017).

\bibitem{FV95} M. Ferconi and G. Vignale, Phys. Rev. B \textbf{52}, 16357 (1995);
 O. Heinonen, M. I. Lubin, and M. D. Johnson, Phys. Rev. Lett. \textbf{75}, 4110 (1995);
 J. Zhao, M. Thakurathi, M. Jain, D. Sen, and J. K. Jain, Phys. Rev. Lett. \textbf{118}, 196802 (2017).

\bibitem{note} The dimensionless $\nu$ is defined as $\nu=N/N_{\phi}$, with $N$ being the total number of electrons, and
$N_{\phi}=AB/\Phi_0$ the Landau level degeneracy. Here $A$ is the area of the sample in the $x-y$ plane ($A^*$ in
$(a_0^*)^2$ units), $B$
is the magnetic field strength, and $\Phi_0 = ch/e$ is the magnetic flux number. Note that $\nu$ may be
written as the ratio between two quantities with dimensions of 2D densities: $\nu = (N/A)/(B/\Phi_0)$.

\bibitem{HM96} C. B. Hanna and A. H. Macdonald, Phys. Rev. B \textbf{53}, 15981 (1996).

\bibitem{units} For material parameters corresponding to {\it GaAs} as well-acting semiconductor,
$Ha^* \simeq 11.04$ meV, and $a_0^* \simeq 101.5 \; \AA$.

\bibitem{MP16} D. Miravet and C. R. Proetto, Phys. Rev. B
\textbf{94}, 085304 (2016).

\bibitem{PBM92} A. P. Prudnikov, Yu. A. Brychkov, and O. I. Marichev, in {\it Integrals and Series, Volume 2, Special Functions}
(Gordon and Breach,New York,1986). See Eq. 2.19.14.15 in page 478.

\bibitem{AS} M. Abramowitz and I. A. Stegun in {\it Handbook of Mathematical Functions} (Dover,New York,1972). 

\bibitem{td} For material parameters corresponding to {\it GaAs} as well-acting semiconductor,
$r_s=2.5$ corresponds to a 2D density equal to  $4.9\times 10^{10}$/cm$^2$, which is a typical
density for the 2D semiconductor systems addressed in this work.

\bibitem{note1} Note that formally this replacement may be motivated by using the relation $n_{\sigma}=\nu_{\sigma}N_{\phi}$.

\bibitem{constant} As stated below Eq. (\ref{DefS2}), for $[\nu_{\sigma}]=0$, 
$S_1^{\nu_{\sigma}}(0)=\nu_{\sigma}^2 I_0^0(0)$,
and $S_2^{\nu_{\sigma}}(0)=\nu_{\sigma} I_0^0(0)$, leading to $C(r_s,0<\nu_{\sigma}<1)=0$.

\bibitem{RHP15} S. Rigamonti,C. M. Horowitz and C. R. Proetto, Phys. Rev. B \textbf{92}, 235145 (2015).

\bibitem{nuinteger} Note that at zero-temperature the EE potential is non-defined for integer $\nu$, 
because of the discontinuities.

\bibitem{HPR97} C. Horowitz, C. R. Proetto, and S. Rigamonti, Phys. Rev. Lett. \textbf{97}, 026806 (2006);
S. Rigamonti and C. R. Proetto, Phys. Rev. B \textbf{73}, 235319 (2006).

\bibitem{HMD00} T. Jungwirth and A. H. MacDonald, Phys. Rev. B \textbf{63}, 035305 (2000).

\bibitem{WMGRBP09} S. Wiedmann, N. C. Mamani, G. M. Gusev, O. E. Raichev, A. K. Bakarov, and J. C. Portal,
 Phys. Rev. B \textbf{80}, 245306 (2009).

\bibitem{MPB16} D. Miravet, C. R. Proetto and P. G. Bolcatto, Phys. Rev. B
\textbf{93}, 085305 (2016). 
 


 




\end{thebibliography}
\end{document}